\documentclass{elsart}
\usepackage{graphicx}
\usepackage{amsmath}
\usepackage{amssymb}
\begin{document}
\begin{frontmatter}
\title{Spectroscopic fingerprints of the frustrated magnetic order 
in $Li_2VOSiO_4$: a $t$$-$$J$ model study}
\author[add1]{I. J. Hamad},
\author[add1]{A. E. Trumper}, 
\author[add1]{L O. Manuel\thanksref{thank1}}
\address[add1]{Instituto de F\'{\i}sica Rosario (CONICET) and
Universidad Nacional de Rosario,
Boulevard 27 de Febrero 210 bis, (2000) Rosario, Argentina\\}
\thanks[thank1]{Corresponding author. Fax: +54-341-4821772. E-mail: manuel@ifir.edu.ar}

\begin{abstract}
We have analyzed theoretically the photoemission spectra of the insulating compound $Li_2VOSiO_4$. 
Recently, this compound has been proposed as the first experimental realization of the 
frustrated $J_1\!-\!J_2$ Heisenberg model. Although it is well known that 
$Li_2VOSiO_4$ is magnetically ordered in a collinear arrangement below $T_N=2.8 K$, there is 
some controversy about the coexistence of two collinear phases above $T_N.$ Using a 
generalized $t\!-\!J$ model we have obtained a complex spectral structure that can be traced 
back to the underlying collinear magnetic structures. We discuss the possibility to use 
ARPES experiments as a way to discern among the different scenarios proposed in the literature.
\end{abstract}

\end{frontmatter}

\begin{figure}[ht]
\begin{center}
\vspace*{0.cm}
\includegraphics*[width=0.25\textwidth]{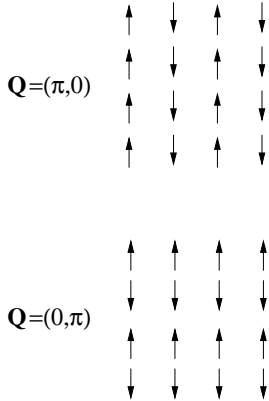}
\caption{Collinear ordered ground states of the $J_1\!-\!J_2$ model
for $J_2/J_1 \ge 0.55$.}
\label{fig1}
\end{center}
\end{figure}

The Heisenberg model on a square lattice with antiferromagnetic (AF) exchange interactions to
first ($J_1$) and second nearest neighbors ($J_2$) is the so-called $J_1\!-\!J_2$ model. It has
all the necessary ingredients to show a rich $T=0$ phase diagram: frustrating exchange
interactions, low spin S=1/2, and low dimensionality. For $J_2/J_1 \ge 0.55$ the ground state
is a twofold degenerate collinear phase (see Fig. \ref{fig1}), with nearest neighboring spins
aligned ferromagnetically along the $x$ direction and antiferromagnetically along the
$y$ direction, or vice versa, characterized by the magnetic wave vector ${\bf Q}=(0,\pi)$ or
$(\pi,0).$ So, the $90^{\circ}$ rotational symmetry of the square lattice is broken.
Even if in 2D models with continous symmetry there cannot be a phase transition at
finite temperature, Chandra, Coleman and Larkin (CCL)\cite{chandra90} argued that the twofold
degeneracy of the collinear ground state could generate a finite temperature Ising-like phase
transition, since now there is a discrete symmetry that can be broken, although 
preserving the spin rotational symmetry. Recent series expansion calculations\cite{singh03}, 
however, do not find evidence for such a finite $T$ transition.
On the other hand, it has been proposed\cite{becca02} another mechanism in which the coupling 
of spins with the lattice is taken into account and a finite temperature Peierls-like transition 
may remove the twofold magnetic degeneracy.

These theoretical issues can be explored in the quasi-2D vanadium oxide $Li_2VOSiO_4,$ which
has been found to be good experimental realizations of the $J_1\!-\!J_2$ model\cite{melzi00}.
In $Li_2VOSiO_4$, the $S=1/2$ can be assumed localized at $V^{4+} (3d^1)$ ions with competing
superexchange interactions to first and second neighbors on a square lattice. NMR\cite{melzi00}, neutron 
scattering\cite{bombardi04}, resonant x-ray scattering\cite{bombardi04}, and magnetization 
measurements\cite{melzi00} indicate that a collinear structure is established below $T_N=2.8 K.$
It should be noticed that the collinear transition at $T_N$ is triggered by the diverging 2D spin 
correlations present in the vanadium layers. Consequently, the $J_1\!-\!J_2$ model is the
appropriate one for the analysis of the magnetic properties of $Li_2VOSiO_4$\cite{melzi00},
though the finite N\'eel temperature is due to 3D residual interactions.
Above $T_N$ the situation is less clear. It has been suggested by NMR experiments\cite{melzi00} 
that just above the N\'eel temperature the degeneracy between the two
possible collinear phases is relieved by a structural distorsion. 
Furthermore, it has been pointed out\cite{carretta02} that magnetic domains of both kind of 
collinear phases coexist for $T_N < T < J_1+J_2.$ This coexistence has been deduced quite 
indirectly from the very low spin dynamics observed in NMR measurements.
However, neutron scattering difraction experiments\cite{bombardi04} do not reveal any structural
phase transition. So, one question that remains to be answered is 
whether the rupture of the discrete degeneracy between the two collinear phases occurs at 
$T_N$ or at a higher temperature $T\sim J_1+J_2$, driven by a purely magnetic mechanism\cite{chandra90}.

Consequently, it is worth to look for an alternative experiment able to detect different magnetic
structures. For this reason, we propose that angle-resolved photoemission experiments (ARPES), in which 
the induced photohole probes the underlying magnetic structure, can give relevant information that 
can be contrasted with the already existing experimental data. In general, photoemission experiments 
are not used to extract magnetic information of a compound
since the spectrum is not able to distinguish between short and 
true long range order. However, it has been recently observed\cite{trumper04} that the photohole 
propagates at different energy scales that are closely related to the 
underlying magnetic correlations, so that ARPES experiments could
be used to discern among different magnetic structures, even short ranged ones. In particular, for
magnetic structures where ferro and AF correlations coexist, we 
have found\cite{trumper04} a many-body quasiparticle excitation that results from the 
coherent coupling of the hole with the AF magnon excitations, while the ferromagnetic
component gives rise to a free-like hole motion at higher energy.

In the present article, we compute hole spectral functions using 
a generalized $t\!-\!J$ model appropriate for $Li_2VOSiO_4.$ Using realistic 
parameters\cite{rosner02} we have found that the photoemission spectra clearly show different 
features when both collinear phases coexist, or not. This result could be helpful in 
elucidating the controversy about the magnetic properties of 
$Li_2VOSiO_4$ in the regime $T_N < T < J_1+J_2.$

To compute the spectral function corresponding to a photohole induced in a photoemission
experiment for $Li_2VOSiO_4,$ we use a generalized $t\!-\!J$ 
model that naturally takes into account the dynamics of a hole injected in the $J_1\!-\!J_2$ model. This 
generalized $t\!-\!J$ model can be written as
\begin{eqnarray}
{H}&=&-t_1\sum_{\langle {i},{j} \rangle 
}(\hat{c}^{\dagger}_{i,\sigma}\hat{c}_{j,\sigma}\!+\!h.c.)-
t_2\sum_{\langle {i},{k} \rangle 
}(\hat{c}^{\dagger}_{i,\sigma}\hat{c}_{k,\sigma}\!+\!h.c.)+\nonumber 
\\
&& +J_1\sum_{\langle {i},{j} \rangle }{{\bf {S}}}_{i} \cdot {{\bf 
{S}}}_{j} + +J_2
\sum_{\langle {i},{k} \rangle }{{\bf {S}}}_{i} \cdot {{\bf 
{S}}}_{k} ,
\label{tj}
\end{eqnarray}
\noindent where $t_1$ ($t_2$) and $J_1$ ($J_2$) correspond to the 
hopping and the exchange
integrals, respectively, to first (second) neighbors. The 
importance of hopping and exchange
interactions to second neighbors is evidenced by the existence of 
a collinear magnetic order below
$T_N$\cite{melzi00} and it is theoretically supported by LDA 
calculations\cite{rosner02}.
In order to calculate the spectral function we use the spinless 
fermion representation\cite{kane89}.
This is a standard procedure that leads to an effective 
Hamiltonian that explicitly includes
the coupling between the hole and the magnon excitations of the 
frustrated magnetic order.
The effective Hamiltonian can be written as:
\begin{eqnarray}
H_{ef}& = & \frac{1}{N}\sum_{k}\epsilon_{k}h_{k}^{\dagger}h_{k}+\frac{1}{\sqrt{N}}
\sum_{kk'}M{}_{kk'}h_{k}^{\dagger}h_{k}\alpha_{k-k'}+\nonumber\\
& & +\sum_{k}\omega_{k}\alpha_{k}^{\dagger}\alpha_{k}  \label{Hef} 
\end{eqnarray}
where the operator $h_{\bf k}^{\dagger}$ ($\alpha^{\dagger}_{\bf 
k}$) creates a hole (magnon)
with momentum ${\bf k}$. $\epsilon_{\bf k}=2t_1\cos k_y$ is the 
bare hole energy,
$M_{\bf kk'}=\beta_{\bf k'}v_{{\bf k-k'}}-\beta_{{\bf k}}u_{{\bf 
k-k'}}$ is the hole-magnon
vertex interaction, where $\beta_{\bf k}=2t_1\sin k_x+4t_2\sin 
k_x\cos k_y,$ and $u_{\bf k},$ $v_{\bf k}$
are the usual Bogoliubov coefficients. $\omega_{\bf k}$ is the 
spin wave dispersion in a collinear
order\cite{chandra90}. The expressions for $\epsilon_{\bf k}$ and 
$\beta_{\bf k}$ have been given for
the collinear order $(\pi,0),$ similar expressions hold for 
$(0,\pi)$.
Treating $H_{ef}$ within the self-consistent Born approximation 
(SCBA)\cite{kane89},
a self-consistent equation for the self energy $\Sigma_{\bf 
k}(\omega),$ at $T=0$, can be derived
$$\Sigma_{\bf k}(\omega )=\frac{1}{N}\sum_{\bf q}
\mid M_{\bf kq}\mid^{2} G_{{\bf k}-{\bf q}}
(\omega-\omega_{\bf q}),$$
where the Green function is defined as $G_{\bf k}(\omega)=(\omega 
-\epsilon_{\bf k}
-\Sigma_{\bf k}(\omega))^{-1}.$ Once the self energy is 
calculated, the spectral function can
be computed as $A_{\bf k}(\omega)=-\frac{1}{\pi} Im \;G_{\bf 
k}(\omega)$.

Theoretically, at finite low temperatures, $T<J_1+J_2,$ it is 
expected a considerable 2D
magnetic correlation length $\xi\sim \exp (2\pi \rho_s/T),$ where 
$\rho_s$ is the spin
stiffness. Therefore, the low energy excitations at finite 
temperatures
still resemble the magnon excitations of the ordered ground 
state. As the spectral function
is insensitive to the long range order, a $T=0$ calculation that 
considers the coupling
between the photohole and magnons will be valid for 
$T_N<T<J_1+J_2.$

Regarding the coexistence of both collinear phases in the range 
$T_N<T<J_1+J_2$, it has been
suggested\cite{carretta02} that the size of the magnetic domains 
is proportional to the correlation
length $\xi$. If we assume that the domain sizes are sufficiently 
large, it is a reasonable
approximation to consider that the photohole dynamics is only 
affected by the magnetic domain
where it was induced. As a consequence, the expected spectral 
function in the case of coexistence
of both orders will be the incoherent sum of the contribution of 
each order, that is,
$A_{\bf k}(\omega)=\frac{1}{2}[A^{(\pi,0)}_{\bf 
k}(\omega)+A^{(0,\pi)}_{\bf k}(\omega)],$
as we assume that the photoelectron is extracted with equal 
probability from both kind of magnetic domains.

\begin{figure}[ht]
\begin{center}
\vspace*{0.cm}
\includegraphics*[width=0.50\textwidth]{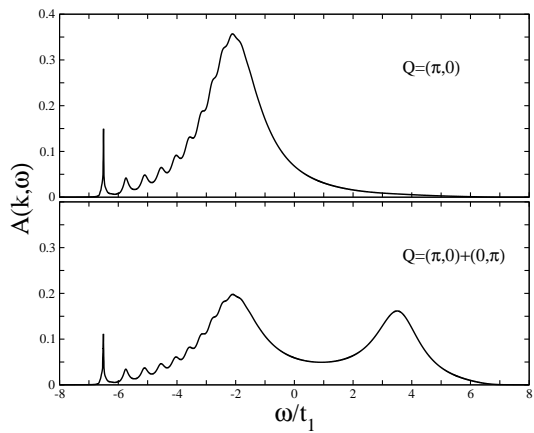}
\caption{
Single-hole spectral function $A({\bf k},\omega)$ for 
${\bf k}=(\pi,0)$ and
realistic parameters for $Li_2VOSiO_4$ (see text).  Upper panel: 
spectral function for the hole
propagating in the $(\pi,0)$ collinear order. Lower panel: 
spectral function for the hole
propagating in a magnetic background of coexistent $(\pi,0)$ and 
$(0,\pi)$ orders.}
\label{fig2}
\end{center}
\end{figure}

To obtain the spectroscopic fingerprints corresponding to the 
coexistence, or not, of different
magnetic domains for $T_N < T < J_1+J_2,$ we have evaluated the 
spectral functions for each case
using the LDA estimated microscopic parameters for $Li_2VOSiO_4$: 
$t_1= -8.5meV,$ $t_2=-29.1meV,$
$J_1=0.83 K$ and $J_2=9.81 K$\cite{rosner02}. In the upper pannel 
of Figure \ref{fig2} it is
shown the spectral function for a hole with momentum ${\bf 
k}=(\pi,0)$ propagating in the
collinear order characterized by the wave vector ${\bf 
Q}=(\pi,0).$ This spectral function has a
finite lifetime resonance located near $\omega=2t_1,$ that can be 
associated with the hole
propagation along the ferromagnetic chains in the $y$ direction. 
At the bottom of the spectral
function there is a weak quasiparticle peak, whose origin can be 
traced back to the coherent
coupling of the hole with the magnons\cite{kane89,trumper04}. We 
have observed that the broad
resonance disperses along the $k_y$ direction like $\epsilon_{\bf 
k},$ with a bandwidth
$\sim 4t_1,$ while the quasiparticle peak exhibits a much smaller 
dispersion of order $2J_2.$

As we mentioned above, for the case in which $(\pi,0)$ and 
$(0,\pi)$ orders coexist
(bottom panel of Fig. \ref{fig2}) we have computed the spectral 
function as the incoherent sum of the contribution of each order. Due to the 
presence of the two kinds of
collinear arrangements, the selected photohole with momentum 
${\bf k}=(\pi,0)$, propagates
preferently at two different energy scales, separated by an 
energy $\sim 6t_1,$ and which are
related to the propagation along the ferromagnetic chains of each 
magnetic background. The
differences between the spectral function for $(\pi,0)$, $A({\bf 
k},\omega)^{(\pi,0)},$ and the
incoherent sum are most clearly observable for momenta close to 
$(\pi,0)$ or $(0,\pi)$, while
along the Brillouin zone diagonal, $\Gamma \to (\pi,\pi),$ the 
spectral functions are equal
for both orders. It should be noticed that, in the case of 
coexistence of phases, the
discrete broken lattice symmetry is restored and, consequently, 
the spectral functions along
the high-symmetry directions $\Gamma \to (\pi,0)$ and $\Gamma \to 
(0,\pi)$ coincide.

Another problem that still remains unsolved in $Li_2VOSiO_4$ is 
the lack of an accurate estimation
of the degree of frustration $J_2/J_1$.
While there is a general consensus that $J_1+J_2 \sim 8.2 K$, 
there is no agreement about the
ratio $J_2/J_1$, with proposed values ranging from 
$1.1$\cite{melzi00} to $12$ \cite{rosner02}.
Contrary to what happens with the magnetic
susceptibility and the specific heat\cite{rosner02}, we have 
observed that
several observable features of the spectra are sensitive to the 
Hamiltonian parameters.
For instance, within our scheme, the first moment of the spectrum 
for momentum $\bf k$,
that is, its center of gravity $<\epsilon>_{\bf k}=\int \omega 
A({\bf k},\omega) d\omega,$
is equal to the hole free hopping energy $\epsilon_{\bf k}$, with 
a bandwidth $W=4t_1.$
The second moment of the spectrum
$<\Delta \epsilon>_{\bf k}= \sqrt{\int (\omega-<\epsilon>_{{\bf 
k}})^2 A(\bf k,\omega)}
d\omega$ characterizes its energy extension and it is equal to 
the average hole-magnon
vertex interaction $\frac{1}{N}\sum_{\bf q}|M_{{\bf k},{\bf 
q}}|^2\sim 3t_1+5t_2.$
Finally, we have found that the weak quasiparticle signal
at the threshold of the spectra disperses with a bandwidth $\sim 
2J_2,$ because of the
predominance of AF correlations between second nearest neighbors.
As we can see, all the above spectral features are intimately 
related to the presence of
collinear magnetic order, and, in principle, they can be read off 
directly from ARPES
experiments, allowing an experimental estimation of the 
microscopic parameters.

In conclusion, we have presented a theoretical calculation of the 
photoemission spectra
for $Li_2VOSiO_4,$ the first experimental realization of the 
$J_1\!-\!J_2$ model. We have used a
generalized $t\!-\!J$ model that takes proper
account of the photohole dynamics in a collinear magnetic 
background. We have shown that
it is possible to extract spectral fingerprints of the underlying 
magnetic structures,
directly associated with the ferro and AF components of the 
magnetic order.
Therefore, we suggest that ARPES experiments would give further 
insight into open
questions about this compound, complementing already available 
data.
Of particular theoretical interest is the coexistence or not of 
the two collinear
orders, $(\pi,0)$ and $(0,\pi)$, in the temperature range $T_N < 
T < J_1+J_2.$
Our results point out that ARPES spectra can clearly distinguish 
between both
situations, because the presence and dispersion of broad 
resonances in
the spectral functions depend crucially on the underlying 
magnetic structures.
We have shown that these broad resonances correspond to the 
photohole propagation
along the ferromagnetic chains of the collinear domains.
Moreover, we have found that several spectral features, like the 
quasiparticle and
the center of gravity dispersions, would allow an experimental 
determination of the
hopping and exchange energies.

This work was supported by the ANPCYT under grant PICT2004 
N${\circ}$ 25724.
I.J.H. thanks Fundaci\'on J. Prats for partial support.


\begin{thebibliography}{99}

\bibitem{misguich03} See, for example, G. Misguich, C. Lhuillier,
cond-mat/0310405 and reference therein.
\bibitem{chandra90} P. Chandra, P. Coleman, A.I. Larkin, Phys. Rev. Lett. 64 (1990) 88.
\bibitem{singh03} R.R.P. Singh, W. Zheng, J. Oitmaa, et al., Phys. Rev. Lett. 91 (2003) 017201.
\bibitem{becca02} F. Becca, F. Mila, Phys. Rev. Lett. 89 (2002) 037204.
\bibitem{melzi00} R. Melzi, P. Carretta, A. Lascialfari, et al.,
Phys. Rev. Lett. 85 (2000) 1318;\\
R. Melzi, S. Aldrovandi, F. Teodoldi, et al., Phys. Rev. B 64 
(2001) 024409.
\bibitem{bombardi04} A. Bombardi, J. Rodriguez-Carvajal, S. Di 
Matteo, et al.,
Phys. Rev. Lett. 93 (2004) 027202.
\bibitem{carretta02} P. Carretta, R. Melzi, N. Papinutto, et al.,
Phys. Rev. Lett. 88 (2002) 047601;\\
P. Carretta, N. Papinutto, R. Melzi, et al.,
J. Phys. Condens. Matter 16 (2004) S894.
\bibitem{rosner02} H. Rosner, R.R.P. Singh, W.H. Zheng, et al.,
Phys. Rev. Lett. 88 (2002) 186405; Phys. Rev. B 67 (2003) 014416.
\bibitem{trumper04} A.E. Trumper, C.J. Gazza, L.O. Manuel, Phys. 
Rev. B 69 (2004) 184407;\\
I.J. Hamad, L.O. Manuel, G. Martinez, et al., Phys. Rev. B 
(2006).
\bibitem{kane89} C.L. Kane, P.A. Lee, N. Read, Phys. Rev. B 39 (1989) 6880;\\
G. Martinez, P. Horsch, Phys. Rev. B 44 (1991) 317.\\
\end{thebibliography}
\end{document}